\documentclass[12pt]{article}
\usepackage{amsmath}
\usepackage{graphics}
\usepackage{latexsym}

\usepackage [ansinew]{inputenc}
\usepackage {hyperref}
\begin{document}

\vskip 2cm
\title{Spinning Pulsating String Solitons in $AdS_5 \times S^ 5$}
\author{\\
A. Khan${}^{|}$ and A.L. Larsen${}^{*}$}
\maketitle
\noindent
{\em Physics Department, University of Southern Denmark,
Campusvej 55, 5230 Odense M,
Denmark}

\vskip 8cm
\noindent
$^{|}$Electronic address: patricio@fysik.sdu.dk\\
$^{*}$Electronic address: all@fysik.sdu.dk
\newpage
\begin{abstract}
\baselineskip=1.5em
\hspace*{-6mm}   We point out the existence of some simple string solitons in
 $AdS_5 \times S^ 5$, which at the same time are spinning in  $AdS_5 $ and
pulsating in  $ S^ 5$,
 or vice-versa. This introduces an additional arbitrary constant into the
scaling relations between energy and spin
 or R-charge. The arbitrary constant is not an angular momentum, but
 can be related to the amplitude of the pulsation.     We discuss the
 solutions in detail and consider the scaling relations.       Pulsating
multi spin or multi R-charge solutions can
 also be constructed.

\end{abstract}

\newpage
\newpage
\section{Introduction}
\setcounter{equation}{0}
Recent progress in understanding the conjectured duality
\cite{maldacena,gubser,witten}
between super string theory on $
AdS_5\times S^5$ and
${\cal N}=4$ SU(N)
super Yang-Mills theory in Minkowski space is based on scaling relations
between energy and angular momentum
for straight spinning strings \cite{klebanov,tseytlin,tseytlin2,russo,armoni,
mandal,minahan,
barbon,axenides,alishabiba,buchel,rashkov,ryang}. Circular pulsating
strings have also been analysed in some detail
\cite{klebanov,minahan}.
Even more recently, multi spin and multi R-charge solutions were constructed
\cite{tseytlin3,tseytlin4,tseytlin5} and classified \cite{tseytlin6,tseytlin7}.

The geometry of Anti de Sitter space profoundly alters the
properties of strings, as compared
to Minkowski space. For instance, for a straight spinning string in Anti de
Sitter space \cite{inigo}, the
energy scales with the angular momentum, while in Minkowski space it scales
with the square root of the
angular momentum. Similarly, for a circular  pulsating string in Anti de
Sitter space \cite{de vega}, the energy
scales with the square of the amplitude, while in Minkowski space it scales
with the amplitude (the
amplitude is of course not a coordinate-invariant, but the maximal
circumference is).

Comparison with Yang-Mills computations (see for instance
\cite{gross,georgi,floratos,korchemsky,dolan})
is based on detailed analysis of the subleading terms
in the scaling relations. It is therefore of impotance to search for
general families of strings
for which it is still possible to obtain analytical results.
In the present paper, we  explicitly construct new families of string
solitons in $AdS_5\times S^5$.
Our strings are straight and spinning in one direction but circular and
pulsating in another, and
with a non-trivial coupling between the two. Pulsating multi spin solutions
are also constructed.
In each case, we obtain the explicit solutions in terms of elliptic
functions, analyse the scaling
relations in various limits, and compare with previously known results. We
note in passing that our solutions
fall outside the classification of \cite{bozhilov,bozhilov2}

The paper is organised as follows:
In Section 2, we set our notation and conventions, and derive and
analyse the solutions which are
spinning in $AdS_5$ but pulsating in $S^5$. Section 3 is devoted to the
"opposite" situation, i.e.
pulsating in $AdS_5$ but spinning in $S^5$. In Section 4, we take the
simplest multi spin solution
\cite{tseytlin3} and generalise it to be pulsating in $S^5$. Finally in
Section 5, we
present our conclusions and we suggest some further investigations in these
directions.

\section{ Spinning in $AdS_ 5$ but Pulsating in $S^5$}
\setcounter{equation}{0}
We take the $AdS_5\times S^5$  line-element  in the form
    \begin{eqnarray}
 ds^2&=&-(1+H^2r^2)dt^2+\frac{dr^2}{1+H^2r^2}+r^2(d\beta^2+\sin^2\beta
d\phi^2+\cos^2\beta d\tilde{\phi}^2 )
 \nonumber \\
 &+&  H^{-2}(d\theta^2+\sin^2\theta d\psi^2+
  \cos^2 \theta (d\psi_1^2+\sin^2\psi_1 d\psi_2^2+\cos^2\psi_1 d\psi_3^2 ))
    \end{eqnarray}
where  $H^{-1}$ is the scale of $AdS_5$ and the radius of $S^5$. The 't
Hooft coupling in this notation is
$\lambda =(H^2\alpha')^{-2}$.

The string which is straight and spinning in $AdS_ 5$ but circular and
pulsating in $S^5$, is obtained by the
 ansatz
\begin{equation}
t=t(\tau),\ \ \ r=r(\sigma), \ \ \ \beta=\pi/2,\ \ \  \phi=\omega\tau, \ \ \
\theta=\theta(\tau), \ \ \
\psi=\sigma
\end{equation}
with the remaining coordinates being arbitrary constants.

The $t$-equation is solved by $t=c_0\tau$, and then the $r$ and
$\theta $ equations become
  \begin{equation}
r''-\frac{H^2{r'}^2r}{1+H^2r^2}-H^2c_0^2(1+H^2r^2)r+\omega^2(1+H^2r^2)r=0
\end{equation}
\begin{equation}
\ddot\theta+\sin\theta\cos\theta=0
\end{equation}
while the non-trivial conformal gauge constraint  is
\begin{equation}
\frac{H^2{r'}^2}{1+H^2r^2}+\dot\theta^2-H^2c_0^2(1+H^2r^2)+H^2r^2\omega^2+\sin^2
\theta=0
\label{2.5}
\end{equation}
Notice that Eq.(\ref{2.5}) involves $\tau$ and $\sigma $ derivatives.
However, everything is consistently solved by
            \begin{eqnarray}
r(\sigma
)&=&\frac{c_0\cos\alpha_0}{\sqrt{\omega^2-H^2c_0^2}}
\mbox{cn}\left(\sqrt{\omega^2-H^2c_0^2\sin^2\alpha_0}\
\sigma|\frac{H^2c_0^2
\cos^2\alpha_0}{\omega^2-H^2c_0^2\sin^2\alpha_0}\right)
\label{2.6}
 \end{eqnarray}
\begin{eqnarray}
 \sin\theta(\tau)&=& \left\{ \begin{array}{ll}
 Hc_0\sin\alpha_0{\mbox {sn}}(\tau |H^2c_0^2\sin^2\alpha_0 )\ ,  \ \ &
Hc_0\sin\alpha_0<1   \\
     {\mbox {sn}}(Hc_0\sin\alpha_0\tau |1/(H^2c_0^2 \sin^2\alpha_0))\ ,\ \
& Hc_0\sin\alpha_0>1 \end{array}     \right.
\label{2.7}
\end{eqnarray}
     where $\alpha_0$ is an integration constant.  In the first case of
Eq.(\ref{2.7}) the string oscillates around one
of the poles,
     while in the second case it oscillates between the two poles. In the
limiting  case, $Hc_0\sin\alpha_0=1$, it oscillates
     between a pole and the equator.  To ensure that $r(\sigma )$ is
periodic, we have the condition
        \begin{eqnarray}
2\pi\sqrt{\omega^2-H^2c_0^2\sin^2\alpha_0}&=&
4K\left(
\frac{H^2c_0^2\cos^2\alpha_0}{\omega^2-H^2c_0^2\sin^2\alpha_0}\right)
\label{2.8}
    \end{eqnarray}
      where $K$ is a complete elliptic integral.

       The solution (\ref{2.6})-(\ref{2.7}) is parametrised by $(c_0, \omega,
\alpha _0)$, of which one is fixed by
Eq.(\ref{2.8}).
        It is convenient
     to trade the two remaining parameters for
\begin{eqnarray}
A\equiv Hc_0\sin\alpha _0
\end{eqnarray}
\begin{eqnarray}
B\equiv  \frac{Hc_0\cos\alpha_0}{\sqrt{\omega^2-H^2c_0^2}}
\label{2.10}
\end{eqnarray}
$A$ is the amplitude of $\sin\theta(\tau)$, while $B$ is the extension of
$Hr(\sigma )$.
    Actually, $A$ can only be interpreted as the amplitude as long as  $
Hc_0\sin\alpha _0  \leq 1$,
    but since the solution for $\theta(\tau) $ is defined for any  $
Hc_0\sin\alpha _0 $, we also define  the
    amplitude for any $ Hc_0\sin\alpha_0 $. Now (\ref{2.8})-(\ref{2.10}) leads to
\begin{eqnarray}
\tan\alpha _0=\frac{\pi A\sqrt{1+B^2}}{2BK\left(\frac{B^ 2}{{1+B^2}}\right)}
\end{eqnarray}
It is straightforward to compute the conserved energy and spin (the
R-charge is zero).
In the present parametrisation, they are
\begin{eqnarray}
        E&=&\frac{2\sqrt{1+B^2}}{\pi H\alpha'}
        \sqrt{B^2+\frac{\pi^2A^2(1+B^2)}{4K^2\left(\frac{B^
2}{{1+B^2}}\right)}}\ E\left(\frac{B^2}{1+B^2}\right)
    \end{eqnarray}
        \begin{eqnarray}
 S&=&   \frac{2\sqrt{1+B^2}}{\pi H^2\alpha'}
        \sqrt{1+B^2+\frac{\pi^2A^2(1+B^2)}{4K^2\left(\frac{B^
2}{{1+B^2}}\right)}}
 \left(E\left(\frac{B^2}{1+B^2}\right)-\frac{1}{1+B^2}K\left(\frac{B^2}{1+B^2}\right)\right)
    \end{eqnarray}
Now we can consider the short   strings in $AdS^5$ corresponding to $B<<1$.
We get
    \begin{eqnarray}
     E&\approx&\frac{1}{ H\alpha' }    \sqrt{B^2+A^2} \\
     S&\approx&\frac{1}{2 H^2\alpha' } B^2 \sqrt{1+A^2}
    \end{eqnarray}
    such that
        \begin{eqnarray}
        {\alpha'} E^2 \approx \frac{A^2}{H^2\alpha'}+\frac{2S}{\sqrt{1+A^2} }
    \end{eqnarray}
    For $A\approx 0$, corresponding to small oscillations near one of the
poles of $S^5$, we get
\begin{eqnarray}
\alpha' E^2\approx 2S
 \end{eqnarray}  as in Minkowski space.
    For $A\approx 1$, corresponding to oscillations between a pole and the
equator of $S^ 5$, we get
\begin{eqnarray}
E/H\approx \frac{1}{H^2\alpha'}+\frac{S}{\sqrt{2}}
 \end{eqnarray}
  while for $A>>1$, corresponding to high frequency oscillations   between
the poles of $S^ 5$, we get
\begin{eqnarray}
E/H\approx \frac{A}{H^2\alpha'}+\frac{S}{A^2}
 \end{eqnarray}
 and the energy is now completely dominated by the contribution from the
oscillations.

 For $B>>1$, corresponding to long strings in $AdS_5$, we get
    \begin{eqnarray}
     E&\approx &\frac{2}{ \pi
H\alpha'}\left(B^2+\frac{1}{2}\log{B}\right)\sqrt{1+\frac{\pi^2A^2}{
4\log^2{B}}}\\
     S&\approx &\frac{2}{\pi H^2\alpha' }
\left(B^2-\frac{1}{2}\log{B}\right)\sqrt{1+\frac{\pi^2A^2}{ 4\log^2{B}}}
    \end{eqnarray}
    such that
    \begin{eqnarray}
E/H-S\approx \frac{1}{\pi H^2\alpha' }  \sqrt{4\log^2 B+\pi^2A^2}
    \end{eqnarray}            where $B=B(S,A)$ is the solution of
    \begin{eqnarray}
     S&\approx &\frac{2B^2}{\pi H^2\alpha' }
\sqrt{1+\frac{\pi^2A^2}{4\log^2{B}}}
\label{2.23}
\end{eqnarray}
If the logarithm dominates over $A$ in Eq.(\ref{2.23}), we have
 \begin{eqnarray}
    B^2\approx\frac{\pi H^ 2\alpha' S} {2}
        \end{eqnarray}
        such that
            \begin{eqnarray}
E/H-S
&\approx   &    \frac{1}{\pi H^2\alpha' }   \log(\frac{\pi H^ 2\alpha' S}{2})+
\frac{\pi A^ 2}{2H^2\alpha'  \log(\frac{\pi H^ 2\alpha' S}{2})}
\label{2.25}
 \end{eqnarray}
    This formula is valid for arbitrary $A$, provided that $
A<<\log(H^2\alpha' S)  $,
     in particular, it holds for $A\approx 0$ and $A\approx 1$.  If $ A$
dominates over the logarithm in
Eq.(\ref{2.23}), we
     have instead
 \begin{eqnarray}
\frac{  B^2}{\log B}\approx\frac{ H^ 2\alpha' S} {A}
        \end{eqnarray}
        such that
            \begin{eqnarray}
E/H-S
  &\approx&     \frac{A}{H^2\alpha' }     +
    \frac{1}{2\pi^ 2 AH^2\alpha' }  \log^2 \frac{A}{ H^ 2\alpha' S}
\label{2.27}
 \end{eqnarray}
    which holds for           $ A>>\log(H^2\alpha' S)  $.

    It is interesting that (\ref{2.25}) and (\ref{2.27}) are formally quite
similar to the results obtained by Russo
\cite{russo} if we replace the amplitude $A$ with the shifted R-charge $H^
2\alpha' J+2/\pi$. But the physics is of
course
    completely different here.

\section{ Spinning in $S^5$ but Pulsating in $AdS_5$}
\setcounter{equation}{0}
We now consider a  string which is spinning in $S^ 5$ but circular and
pulsating in $AdS_5$. It is obtained by the
 ansatz
\begin{equation}
t=t(\tau),\ \ \ r=r(\tau), \ \ \  \beta=\pi/2,\ \ \   \phi=\sigma, \ \ \
\theta=\theta(\sigma), \ \ \
\psi=\nu\tau
\end{equation}
with the remaining coordinates being arbitrary constants.

The $t$-equation can be integrated to
  \begin{eqnarray}
\dot t=\frac{c_0}{1+H^2r^2}
\label{3.2}
\end{eqnarray}
and then the $r$ and $\theta $ equations become
  \begin{equation}
\ddot r-\frac{H^2\dot
r^2 r}{1+H^2r^2}+\frac{H^2c_0^2r}{1+H^2r^2}+(1+H^2r^2)r=0
\end{equation}
\begin{equation}
\theta''+\nu^2\sin\theta\cos\theta=0
\end{equation}
while the non-trivial conformal gauge constraint  is
\begin{equation}
\frac{H^2\dot
r^2}{1+H^2r^2}+{\theta'}^2-\frac{H^2c_0^2}{1+H^2r^2}+H^2r^2+\nu^2\sin^2\theta=0
\end{equation}
Again, the constraint mixes $\tau$ and $\sigma $ derivatives, but the
solution is easily obtained	  as
\begin{eqnarray}
&r(\tau)=\frac{1}{H\sqrt{2}}\left(\sqrt{(1+H^2c_0^2\sin^2\alpha
_0)^2+4H^2c_0^2\cos^2\alpha _0}-
(1+H^2c_0^2\sin^2\alpha_0)\right)^{1/2}&\nonumber\\
&\times\mbox{cn}\left(((1+H^2c_0^2\sin^2\alpha _0)^2+4H^2c_0^2\cos^2\alpha
_0)^{1/4}\tau
|m\right) &
\end{eqnarray}
\begin{equation}
\sin\theta(\sigma )=
\frac{Hc_0\sin\alpha_0}{\nu}{\mbox {sn}}\left(\nu\sigma
|\frac{H^2c_0^2\sin^2\alpha_0}{\nu^2}\right )
\label{3.7}
\end{equation}
where the elliptic parameter of the $r$-expression is
	\begin{equation}
	m=\frac{1}{2}-\frac{1+H^2c_0^2\sin^2\alpha_0}{2\sqrt{(1+H^2c_0^2\sin^2\alpha_0)
^2+4H^2c_0^2\cos^2\alpha_0}}
\end{equation}
Eq.(\ref{3.2}) can now be integrated in terms of an elliptic integral of the
third kind, but we shall not need the explicit
expression. The solution (\ref{3.7}) is valid for $Hc_0\sin\alpha_0\leq\nu$.
There is another solution for
  $Hc_0\sin\alpha_0>\nu$ but it is believed to be unstable \cite{klebanov}
so we shall not consider it here.
  To ensure periodicity of $\theta (\sigma )$, we have the condition

        \begin{eqnarray}
2\pi\nu&=&4K\left(\frac{H^2c_0^2\sin^2\alpha_0  }{\nu^2}\right)
    \end{eqnarray}
     As in the previous section, it is convenient to trade the two
remaining parameters (say $c_0$ and $\alpha _0$)
     for two new ones
\begin{equation}
A=\frac{1}{\sqrt{2}}\left(\sqrt{(1+H^2c_0^2\sin^2\alpha
_0)^2+4H^2c_0^2\cos^2\alpha _0}-
         (1+H^2c_0^2\sin^2\alpha _0)\right)^{1/2}
\end{equation}
    \begin{eqnarray}
 B=\frac{Hc_0\sin\alpha_0}{\nu}
 \end{eqnarray}
 such that $A$ is the amplitude of $Hr(\tau)$ while $B$ is the extension of
$\sin\theta(\sigma)$. As noted
 in the introduction, the amplitude is coordinate dependent. A coordinate
invariant measure of the oscillations is
 given by the maximal circumference. This is precisely our $A$, up to a
factor of $2\pi$.

 A straightforward computation gives the energy and R-charge  in this
parametrisation (the spin is zero)
    \begin{eqnarray}
 E =\frac{1}{H\alpha'}\sqrt{A^2\left(A^2+1+\frac{4B^2}{\pi^2}K^2(B^2)\right)
 +\frac{4B^2}{\pi^2}K^2(B^2)}
 \end{eqnarray}
    \begin{eqnarray}
    J= \frac{2}{\pi H^2\alpha' }\left( K(B^2)-E(B^2)\right )
    \end{eqnarray}
    First consider short strings in $S^5$ corresponding to $B<<1$
    \begin{eqnarray}
 E \approx \frac{1}{H\alpha'}\sqrt{A^2\left(A^2+1+B^2 \right)
 +B^2}
 \end{eqnarray}
    \begin{eqnarray}
    J\approx \frac{B^2}{2 H^2\alpha'  }
     \end{eqnarray}
     such that
\begin{eqnarray}
 E \approx \frac{1}{H\alpha'}\sqrt{A^2\left(A^2+1+2 H^2\alpha' J \right)
+2 H^2\alpha' J  }
 \end{eqnarray}
 For $A\approx 0$ we get
\begin{eqnarray}
\alpha' E^2\approx 2J
 \end{eqnarray}
 which is like in Minkowski space.  For $A>>1$ we get
 \begin{eqnarray}
 E/H\approx   \frac{A^ 2}{H^2\alpha'} +
 \frac{1}{ 2H^2\alpha'}   +J
 \end{eqnarray}
 such that the energy is completely dominated by the pulsation, $E\sim A^ 2$.

    Now consider long strings in $S^5$ ('long' meaning extending almost
down to the equator) corresponding to $B\approx
1$
    \begin{eqnarray}
  E \approx
\frac{1}{H\alpha'}\sqrt{A^2\left(A^2+1+\frac{1}{\pi^2}\log^2\frac{16}{1-B^2}\right)
 +\frac{1}{\pi^2}\log^2\frac{16}{1-B^2} }
 \end{eqnarray}
    \begin{eqnarray}
      J&\approx &  \frac{2}{\pi
H^2\alpha'}\left(\frac{1}{2}\log\frac{16}{1-B^2}-1\right)
     \end{eqnarray}
     such that
    \begin{eqnarray}
  E \approx \frac{1}{H\alpha'}\sqrt{A^2\left(A^2+1+\frac{1}{\pi^2}(\pi
H^2\alpha' J+2)^2 \right)
 +\frac{1}{\pi^2}(\pi H^2\alpha' J+2)^2}
 \end{eqnarray}
 For $A\approx 0$ we get
    \begin{eqnarray}  E/H-J \approx \frac{2}{\pi  H^2\alpha'} +\frac{JA^2}{2}
 \end{eqnarray}
 For $A>>1$ we must distinguish between different cases. If $A>>H^2\alpha'
J$ we get
    \begin{eqnarray}
  E/H \approx \frac{A^2}{H^2\alpha'}+\frac{1}{ 2}H^2\alpha' J^2
 \end{eqnarray}
 If $A<<H^2\alpha' J$ we get
\begin{eqnarray}
  E/H  \approx AJ+\frac{A^3}{2H^4{\alpha'}^2J}
 \end{eqnarray}
 The scaling relations obtained here are, to our knowledge, completely new.
They generalise the ones
 of Gubser, Klebanov and Polyakov \cite{klebanov} and supplement the ones
of Russo \cite{russo}.

\section{  Pulsating Multi Spin Solutions}
\setcounter{equation}{0}
Multi spin and multi R-charge solutions were recently obtained by Tseytlin
and Frolov
\cite{tseytlin3,tseytlin4,tseytlin5}. Such solutions can
easily be combined with pulsation. Here we take for simplicity the simplest
2 spin solution
in $AdS_5$ and couple it with pulsation in $S^5$. The ansatz is
\begin{equation}
t=c_0\tau,\ \ \ r=r_0={\mbox {const.}}, \ \ \    \beta =\sigma ,\ \ \
\phi=\omega\tau, \ \ \
\tilde{\phi}=\omega\tau,\ \ \
     \theta=\theta(\tau), \ \ \
\psi=\sigma
\end{equation}
with the remaining coordinates being constants.     This is a circular
string in $AdS_5$ spinning in two different
directions. It is also a circle in $S^5$, but pulsating  there.

The       $r$ and $\theta $ equations become
  \begin{equation}
\omega^2=1+H^2c_0^2
\end{equation}
\begin{equation}
\ddot\theta+\sin\theta\cos\theta=0
\end{equation}
while the non-trivial conformal gauge constraint  is
\begin{equation}
\dot\theta^2+\sin^2\theta-H^2c_0^2(1+H^2r_0^2)+H^2r_0^2(1+\omega^2)=0
\end{equation}
The $\theta $ equation and constraint are solved by
\begin{eqnarray}
\sin\theta(\tau)&=& \left\{ \begin{array}{ll}
A{\mbox {sn}}(\tau |A^2 )\ ,  \ \ &A<1  \\
{\mbox {sn}}(A\tau |1/A^2)\ ,\ \ & A>1 \end{array}     \right.
\end{eqnarray}
where
\begin{equation}
A=H\sqrt{c_0^2-2r_0^2}
\end{equation}
     with the same interpretation as in Section 2.    The energy $E$ and
the 2 spins
$S_1=S_2\equiv S$ are easily computed
\begin{eqnarray}\label{7}
E&=&\frac{(1+H^2r_0^2)\sqrt{2H^2r_0^2+A^2}}{H\alpha'}\\
S&=&\frac{r_0^2}{2\alpha'}\sqrt{2r_0^2H^2+A^2+1}\label{4.8}
\end{eqnarray}
For short strings (say $Hr_0<<1$) we get
\begin{eqnarray}\label{7}
S&\approx &\frac{r_0^2}{2\alpha'}\sqrt{A^2+1} \left(1+\frac{H^2r_0^2}{A^2+1}  \right)
\end{eqnarray}
such that
\begin{eqnarray}
 H^2r_0^2\approx       \frac{2H^2\alpha' S}{\sqrt{A^2+1}}-
\frac{4H^4{\alpha'}^2 S^ 2}{(A^2+1)^ 2}
\end{eqnarray}
which inserted into $E$ gives
\begin{eqnarray}
E(S,A)\approx \frac{1}{H\alpha'}\left(1+\frac{2H^2\alpha'S}{\sqrt{A^2+1}}-
\frac{4H^4{\alpha'}
^2S^2}{(A^2+1)^2}\right)\sqrt{\frac{4H^2\alpha'S}{\sqrt{A^2+1}}-
\frac{8H^4{\alpha'} ^2S^2}{(A^2+1)^2}+A^2}
\end{eqnarray}
For $A=0$ we get
\begin{eqnarray}
E\approx \frac{2\sqrt{S}}{\sqrt{\alpha'}}(1+H^2\alpha'S)
\end{eqnarray}
which to leading order is just the Minkowski result $\alpha' E^2=2(2S)$.
For $A=1$ we get
\begin{eqnarray}
  E\approx \frac{1}{H\alpha'}(1+2^{3/2}H^2\alpha'S)
\end{eqnarray}
and for $A>>1$
\begin{eqnarray}
E\approx \frac{1}{H\alpha'}(A+2H^2\alpha'S)
\end{eqnarray}
For long strings (say $Hr_0>>1$) we get
\begin{eqnarray}
E/H-2S
&\approx &\frac{\sqrt{2H^2r_0^2+A^2}}{H^2\alpha'}  -
\frac{H^2r_0^2}{2H^2\alpha'  \sqrt{2H^2r_0^2+A^2}}
\label{4.15}
\end{eqnarray}
Now we have to distinguish between different cases. If $Hr_0>>A$ we get
from (\ref{4.8})
\begin{eqnarray}
S&\approx &\frac{Hr_0^3}{\sqrt{2}\alpha'}\left(1+\frac{A^2+1}{4H^2r_0^2}\right)
\end{eqnarray}
such that
\begin{eqnarray}
Hr_0\approx (\sqrt{2}H^2\alpha' S)^{1/3}-\frac{A^2+1}{12(\sqrt{2}H^2\alpha'
S)^{1/3}  }
\end{eqnarray}
which when inserted into (\ref{4.15}) gives
\begin{eqnarray}
E/H-2 S\approx \frac{3(\sqrt{2}H^2\alpha'S)^{1/3}}{2^{3/2}H^2\alpha'}+
\frac{4A^2-1}{2^{7/2}H^2\alpha'(\sqrt{2}H^2\alpha'S)^{1/3}}
\end{eqnarray}
This result is valid for $H^2\alpha' S>>\{1, \ A^3\}$, and therefore holds
in particular for $A=0$ and $A=1$.
Notice also that the pulsation only gives a contribution to the non-leading
terms, in this limit.
On the other hand, if $Hr_0<<A$ we get from Eq.(\ref{4.8})
\begin{eqnarray}\label{7}
S&\approx &\frac{Ar_0^2}{2\alpha'}\left(1+ \frac{r_0^2H^2}{A^2}\right)
\end{eqnarray}
such that
\begin{eqnarray}\label{7}
H^2r_0^2\approx \frac{2H^2S\alpha'}{A}-\frac{4S^2{\alpha'}^2H^4}{A^4}
\end{eqnarray}
and insertion into (\ref{4.15}) gives
\begin{eqnarray}\label{7}
E/H-2 S\approx \frac{A}{H^2\alpha'}+\frac{S}{A^2}
\end{eqnarray}
which holds for $1<<H^2\alpha' S<<A^3$.  Thus, in this limit, the pulsation
completely changes the
scaling relation.

All our results reduce for $A=0$ to those obtained in \cite{tseytlin3}, but
otherwise they are quite
different.  It is straightforward also to generalise the multi R-charge
solutions of \cite{tseytlin3}
to  include pulsation in $AdS_5$, but we shall not go into the details here.

\section{ Concluding Remarks}
\setcounter{equation}{0}
In conclusion,   we have found several new relatively simple families of
string solitons in $AdS_5\times S^5 $.
They generalise some of the previously known solitons in the sense that
they combine spin and pulsation in
a non-trivial way. For each family, we analysed in detail the scaling
relations between energy and angular
momentum. The scaling relations reduce in a certain limit ($A=0$) to
previously known relations, but are
otherwise quite different.

It would be interesting to find the Yang-Mills  operators corresponding to
the spinning pulsating strings
constructed here. It seems, however, to be a highly non-trivial problem due
to the coupling between spin
and pulsation. The Yang-Mills operators for spinning strings were given in
\cite{klebanov}, and
those for  pulsating strings were suggested in \cite{minahan}, but it is
not clear how to combine them.

Another direction which could be interesting to pursue,
is to consider linearised perturbations around these solutions.
Classically, this could reveal how the pulsation affects the stability
properties of spinning strings, a question
which could be important especially for the multi spin solutions of
\cite{tseytlin3}. At the quantum level,
the perturbations could give a contribution to the scaling relations,
similarly to the results obtained in
 \cite{tseytlin,tseytlin2}.

These problems are currently under investigation.

\end{document}